# Atomic-Scale Tracking Phase Transition Dynamics of Berezinskii-Kosterlitz-Thouless Polar Vortex-Antivortex

Ruixue Zhu[1,2,‡], Sizheng Zheng[3,‡], Xiaomei Li[1,2,7,‡], Tao Wang[1,‡], Congbing Tan[4,*], Tiancheng Yu[1], Zhetong Liu[1], Xinqiang Wang[5,9], Jiangyu Li[6,*], Jie Wang[3,8,*], and Peng Gao[1,2,9,*]

[1]Electron Microscopy Laboratory, School of Physics, Peking University, Beijing 100871, China

[2]International Center for Quantum Materials, School of Physics, Peking University, Beijing 100871, China

[3]Department of Engineering Mechanics, Zhejiang University, Hangzhou 310027, Zhejiang, China

[4]Hunan Provincial Key Laboratory of Intelligent Sensors and Advanced Sensor Materials, School of Physics and Electronics, Hunan University of Science and Technology, Xiangtan 411201, Hunan, China

[5]State Key Laboratory of Artificial Microstructure and Mesoscopic Physics, School of Physics, Peking University, Beijing 100871, China

[6]Guangdong Provincial Key Laboratory of Functional Oxide Materials and Devices, Southern University of Science and Technology, Shenzhen 518055, Guangdong, China

[7]School of Integrated Circuits, East China Normal University, Shanghai 200241, China

[8]Zhejiang Laboratory, Hangzhou 311100, China

[9]Collaborative Innovation Centre of Quantum Matter, Beijing 100871, China

[‡]These authors contributed equally to the work.

*Corresponding authors. Email: cbtan@xtu.edu.cn; lijy@sustech.edu.cn; jw@zju.edu.cn; pgao@pku.edu.cn;

**Abstract:** Particle-like topologies, such as vortex-antivortex (V-AV) pairs, have garnered significant attention in the field of condensed matter. However, the detailed phase transition dynamics of V-AV pairs, as exemplified by self-annihilation, motion, and dissociation, have yet to be verified in real space due to the lack of suitable experimental techniques. Here, we employ polar V-AV pairs as a model system and track their transition pathways at atomic resolution with the aid of in situ (scanning) transmission electron microscopy and phase field simulations. We demonstrate the absence of a Berezinskii-Kosterlitz-Thouless phase transition between the room-temperature quasi-long-range ordered ground phase and the high-temperature disordered phase. Instead, we observe polarization suppression in bound V-AV pairs as the temperature increases. Furthermore, electric fields can promote the vortex and antivortex to approach each other and annihilate near the interface. The elucidated intermediate dynamic behaviors of polar V-AV pairs under thermal- and electrical-fields lay the foundation for their potential applications in electronic devices. Moreover, the dynamic behaviors revealed at atomic scale provide us new insights into understanding topological phase of matter and their topological phase transitions.

Topological phase transitions mediated by nontrivial topologies, such as vortex-antivortex (V-AV) pair, have attracted considerable attention in condensed matter due to their fundamental physics and potential applications in spintronics and electronic memory devices[1]. Of particular interest is the celebrated Berezinskii-Kosterlitz-Thouless (BKT) phase transition[2, 3] in two-dimensional (2D) systems where long-range order and spontaneous symmetry breaking are theoretically prohibited at finite temperatures[4]. BKT phase transition has been extensively studied and demonstrated to exist in various systems, including superconductors[5, 6], superfluid helium[7], Josephson junction arrays[8], and 2D electron gas[9], and is manifested as the emergence of quasi-long-range ordered V-AV pairs and their dissociation. In addition to the characteristic binding-unbinding transition that generates isolated vortices and antivortices[10], spin-type V-AV is expected to exhibit other dynamic behaviors, such as motion and self-annihilation, which are sufficiently supported by theoretical investigations[11, 12].

However, reliable real space experimental evidence for these evolutions of V-AV is scarce, limited by existing characterization techniques. For example, the commonly used Lorentz transmission electron microscopy (LTEM) for studying magnetic topologies suffers from vulnerable phase contrast, and LTEM results are highly sensitive to experimental conditions such as defocus[13-16]. Moreover, due to the weak magnetic signal in quasi-2D magnetic materials, LTEM usually requires a large defocus to record data[17], which prevents the realization of ultrahigh spatial resolution. Extending research into the field of ferroelectrics, several simulation studies have predicted that a BKT phase transition may occur in 2D ferroelectric systems with an increase in temperature, accompanied by a power-law decay of the correlation function[18-20]. Experimentally, Chae et al. used atomic force microscopy to determine, at the micron scale, that the temperature-dependence transition of pairwise polar vortices in layered hexagonal $ErMnO_3$ is not of the BKT type[21]. Kim et al. demonstrated the biased-tip-induced creation and separation of a polar V-AV pair in $BiFeO_3$ film via angle-resolved lateral piezoresponse force microscopy (PFM) at the hundred-nanometer scale[22]. However, PFM contrast is highly sensitive to multiple external factors, such as the coupling of tip and moving cantilever, and it is difficult to distinguish the contribution of ion potential and polarization charge, making data interpretation vulnerable[23-26]. To the best of our knowledge, conclusive real-space evidence for dynamics of BKT-type V-AV pairs is still lacking.

When considering spatial resolution, characterization techniques for ferroelectrics are superior to those for ferromagnets. Various atomic imaging techniques can map subunit-scale polarization with high precision[27-30], making in situ tracking of the topological phase transition dynamics of polar V-AV pairs even more compelling. $PbTiO_3$(PTO)/$SrTiO_3$(STO) superlattice is paradigmatic system for investigating various polar topologies involving flux-closures[31-33], vortex[34, 35], skyrmions[36], and their topological phase transitions under external stimuli[37-44]. By manipulating the layer thickness of PTO/STO superlattice grown on $DyScO_3$ (DSO) substrate, polar V-AV pairs have been demonstrated to exist stably at room temperature, thanks to the delicate tuning of elastic, electrostatic and gradient energies[45-48].

In this context, we consider quasi-long-range ordered polar V-AV pairs in the fabricated quasi-2D $(PTO)_{11}/(STO)_6$ superlattice lamella as the ground state to investigate their possible dynamic behaviors under thermal and electrical stimuli. As schematically shown in Fig. 1a, the terminals of thermal- and electric-field-induced transition are disordered paraelectric phase and long-range-ordered ferroelectric monodomain. Combining atomically resolved in situ spherical aberration-corrected transmission electron microscopy and phase field simulation, we find that polar vortices and antivortices coexist and behave as bound pairs over a wide temperature range, but their polarization magnitude gradually decreases upon heating. No unbinding of polar V-AV pairs occurs between the quasi-long-range ground state and paraelectric state. In addition, electric fields can prompt the motion of vortex cores and antivortex cores, resulting in the decrease of the distance between adjacent V-AV and culminating in their annihilation near the interface. Our findings reveal the dynamic details of polar V-AV pairs at the scale of electric dipole, and provide useful information for their potential applications in future, as well as effective guidance for exploring the BKT phase transition in ferroelectric materials.

**Results**

**Polar V-AV pairs in designed superlattice**

Guided by the V-AV pair phase diagram dependent on the PTO/STO layer thickness[45], $(PTO)_{11}/(STO)_6$ superlattices were grown on DSO substrates using pulsed laser deposition for this study, with embedded $SrRuO_3$ (SRO) layer serving as the bottom electrode (details in the Methods). A typical atomic-scale middle angle annular dark field (MAADF) scanning transmission electron microscopy (STEM) image taken along the [100] zone axis reveals the alternate high-contrast PTO layers and low-contrast STO layers with sharp and coherent interfaces (Fig. 1b). Vortex arrays exist in the PTO layers, as indicated by the diffraction contrast in the TEM dark-field image (Fig. 1c), and weak strain contrast in the MAADF image. A magnified high angle annular dark field (HAADF) image is shown in Fig. 1d, from which we can derive the displacement vectors of Ti relative to the four neighboring cations[27]. The corresponding polar vector map is plotted on the left panel of Fig. 1e, confirming the existence of bound V-AV pairs in the superlattice. Notably, antivortices in the STO layer are sandwiched between pairs of aligned vortices with the same circulation in adjacent PTO layers, in good agreement with the results of phase field simulations depicted in the middle and right panels. To better appreciate polar vector distribution of antivortices, the lengths of all arrows in the right panel have been standardized. The distribution of polar V-AV can also be reflected by the unit-cell-level maps of out-of-plane lattice constant $c$ (Fig. 1f) and in-plane lattice constant $a$ (Fig. 1g). These maps were calculated based on Fig. 1d, which is highly consistent with the simulation results in Fig. 1h and Fig. 1i. Obviously, the $c$-color map in Fig. 1f exhibits a characteristic orange sinusoidal wave pattern in the PTO layers, where the local maximum corresponds to the most strained vortex core, marked by purple balls. Notably, the staggered vortex cores are located at different heights. In addition, the $a$-color map in Fig. 1g can also help us identify the vortex cores, which are located at the vertices of the orange triangles. In the STO layer, the $a$-color map is inhomogeneous and exhibits an orange-colored wave feature, which is more pronounced in the simulation image (Fig. 1i). The antivortex cores are located in the gap of the orange wave, as marked by orange balls. Apart from lattice analysis, we also carried out accurate quantitative measurements of the out-of-plane displacement $dz$ and in-plane displacement $dx$ of Ti cations. The corresponding color magnitude maps are displayed in Fig. 1j and Fig. 1k, and agree well with the simulation results in Fig. 1l and Fig. 1m. Referring to the displacement vectors in Fig. 1e, it is evident that the polarization distribution of the vortex is asymmetric, manifested by unequal in-plane domains and zig-zag vortex cores[30].

**No unbinding behavior of V-AV pairs under thermal field**

Taking the quasi-long-range ordered polar V-AV pairs as the ground state, we further explored their thermal responsive behaviors by performing in situ heating experiments on the specimen using a chip-based approach (details in the Methods). During the heating process, atomically resolved HAADF images were collected at 25 °C, 300 °C and 400 °C (Fig. 2a), followed by Gaussian fitting to determine the atomic column positions for quantitative analysis. The initial lattice constant color maps in Fig. 2b and Fig. 2c show representative sinusoidal waves belonging to V-AV pairs. As the temperature increased, the orange sinusoidal wave patterns in the PTO layers become less pronounced (Fig. 2b), indicating a reduction and homogenization of the out-of-plane strain, and thus the disappearance of polar vortices, exactly as expected from the simulations (supplementary fig. 1). Focusing on the $a$-color map change of STO layer in Fig. 2c, the inhomogeneous in-plane strain gradually becomes uniform, which reflects the thermal erasability of antivortices. The effects of thermal expansion are excluded in both experimental calculations and phase field simulations[49, 50]. On this basis we further performed quantitative statistics of lattice $c$ and lattice $a$ (Fig. 2d, Fig. 2e and supplementary fig. 2), using the standard deviation (std) as a measure. With heating, the stds of both $c$ and $a$ exhibit a decreasing trend, indicating that the dispersion of lattice constants is reduced due to the disappearance of polar structures. As a reference, we conducted quantitative analysis of $(PTO)_7/(STO)_{13}$ superlattice which only contains vortex but no antivortex (supplementary fig. 3). Indeed, the stds of $c$ and $a$ in the STO layer without antivortex were relatively lower than those in the $(PTO)_{11}/(STO)_6$ superlattice containing antivortex.

The polarization change of V-AV pairs with temperature is also intriguing, as reflected by the unit-cell-scale measurements of out-of-plane displacement $dz$ and in-plane displacements *dx* (see Fig. 3a and Fig. 3b). The variation of displacement magnitude profiles, as exemplified by the black dashed frames, quantitatively describes the reduction of ferroelectricity with increasing temperature (Fig. 3c and Fig. 3d). The topological phase transition from V-AV to paraelectric phase occurs at ~ 400 °C. It is worth noting that vortex and antivortex always coexist as bound pairs, rather than annihilation or separation into isolated individuals. When T = 370 °C, the average polarization of antivortex is only ~ 0.43 µC/cm$^2$, it can be considered that its ferroelectricity almost disappears. At this time, the average polarization of vortex is still ~ 4.06 µC/cm$^2$. That is to say, the antivortex will disappear prior to vortex during the heating process, because the bound charge on which the stability of antivortex depends is provided by adjacent vortices[45]. Additionally, the staggered vortex cores at room temperature tend to align horizontally at high temperature, as evident from the simulation results in Fig. 3e and Fig. 3f, which is attributed to the combined effect of increased Landau energy and decreased elastic energy (supplementary fig. 4)[51]. In accordance with phase field simulations, we construct a temperature phase diagram in Fig. 3g, showing the temperature dependence of the average polar displacement |D|, and divide it into three regions: (i) 25 °C-370 °C, V-AV pairs are dominant; (ii) 370 °C-400 °C, antivortex disappears and vortex remains; (iii) above 400 °C, paraelectric phase without any polar structures.

**Approach and annihilation of V-AV pairs under electric field**

An electric field was then applied to polar V-AV pairs through the bottom electrode SRO and the carbon layer deposited on the surface in order to further explore their electrical responsive behaviors. A typical dark-field TEM image series (Fig. 4a) shows alternating bright and dark dotted-like contrast, initially representing vortex arrays, tend to become tilted stripes, indicating a topological phase transition from vortices to trivial a/c domains (supplementary fig. 5)[39, 42]. With further increase of the electric field, the PTO layers display a uniformly polarized state, suggesting the formation of ferroelectric monodomains. Simulations provide a consistent transition process (Fig. 4b). A striking finding is that the cores of vortex and antivortex can move and approach each other under the electric field until they annihilate near the interface.

The corresponding dipole evolution is depicted in Fig. 4c. Furthermore, a phase diagram has been conducted showing the electric-field-dependent evolution of polar V-AV pairs along with their energy changes, as shown in Fig. 4d. Three regions are divided: (i) 0-200 kV/cm, V-AV pairs are dominant; (ii) 200 kV/cm-750 kV/cm, V-AV pairs are broken and a/c domains are dominant; (iii) above 750 kV/cm, c domains are dominant.

**Discussion**

The underlying mechanism of these phenomena is briefly discussed below. In magnetic systems, the interaction between vortex and antivortex composed of spins is always attractive, and they tend to mutually annihilate when approaching each other[12, 52]. Stable bound V-AV pairs are only possible in specific scenarios, such as when they are confined within narrow temperature ranges or certain pining states[53-56]. The dissociation of the V-AV pairs occurs at the BKT transition temperature, resulting in isolated free vortices[5, 57]. However, in our ferroelectric superlattice regime, polar V-AV pairs can form spontaneously under delicate strain and electrostatic boundary conditions and remain stable at room temperature in the absence of external fields. The pinning effect of the interface prevents their self-annihilation, but the electric field can promote vortex and antivortex cores to approach each other and annihilate near the interface. Additionally, no binding-unbinding phenomenon of polar V-AV pairs is observed during heating. A possible reason for this occurrence could be the introduce of the PTO/STO interface, which may provide a pinning potential that significantly perturbing the separation of polar V-AV pairs.

The 2D conditions are a prerequisite for discussing the occurrence of BKT phase transition, according to the previous theorem[1, 2]. In our study, the PTO/STO superlattice films were fabricated into nanometer-thick lamellas using ion beam etching. The sample geometry met the necessary requirements. However, it is worth noting that the strain conditions for our sample differ slightly from those of the as-obtained thin film. Specifically, the original biaxial strain provided by the substrate becomes uniaxial[58], which may lead to polar structure reconstruction and phase transition pathways that differ from those studied using X-ray diffraction techniques[40, 41], particularly in our case where the high-temperature $a_1/a_2$ phase was not observed.

In conclusion, we use room-temperature quasi-long-range ordered polar V-AV pair arrays as the ground state to quantitatively reveal their topological phase transition dynamics under thermal and electric fields with atomic-level spatial resolution. We find that either thermal or electric fields can promote the topological phase transition of polar V-AV pairs, but the detailed intermediate transition process, including self-annihilation, dissociation, motion, etc., are distinct and have not been elucidated. Specifically, no dissociation-like behavior associated with BKT transition was observed during heating until the paraelectric phase was formed. The superlattice interface facilitates the formation of polar V-AV pairs, but prevents the occurrence of BKT phase transition, as well as the self-annihilation of V-AV. However, under the action of electric fields, the cores of polar vortex and antivortex can gradually approach and annihilate near the interface, resulting in the transition from vortex phase to a/c domains. Our in situ work on polar V-AV pairs in quasi-2D ferroelectric superlattice lamella illuminates their missing evolutionary pathways during the topological phase transitions, which provides instructive insights for future experimental exploration of ferroelectric BKT phase transition. Furthermore, the revealed dynamic details of polar V-AV under external stimuli provide essential information for the design of devices based on them.

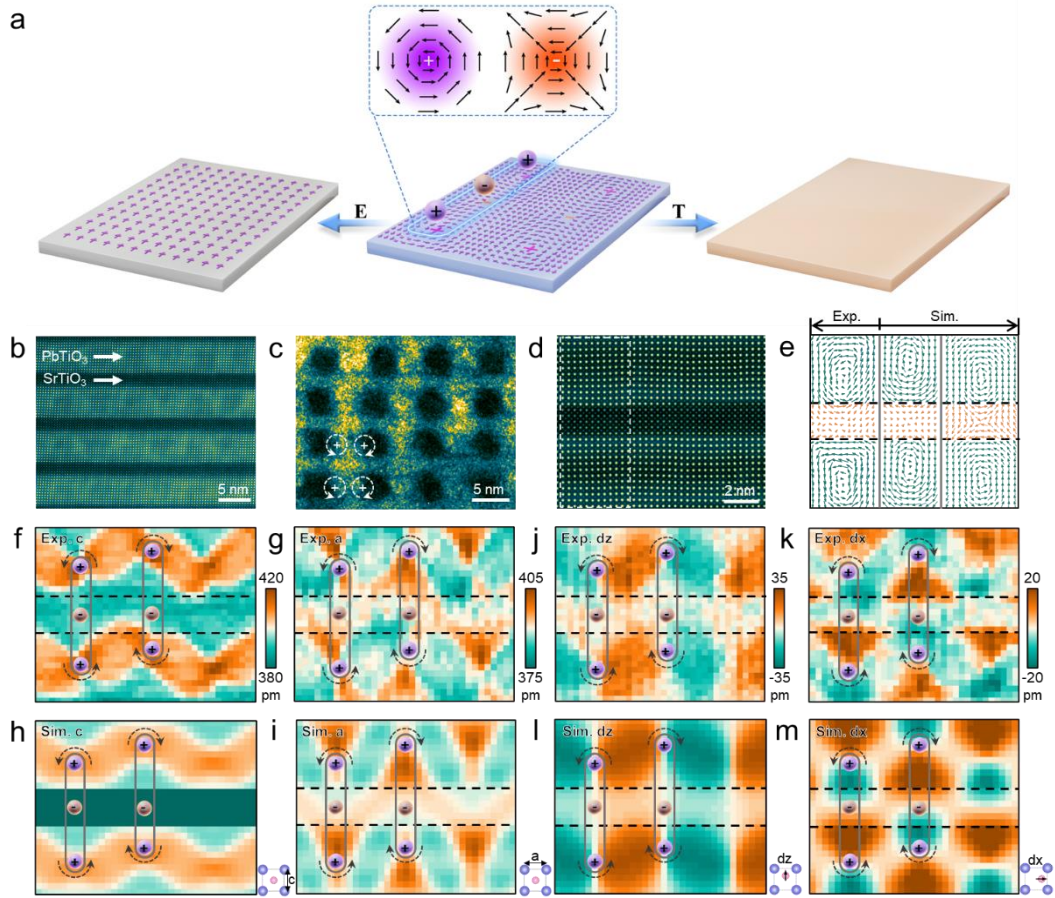

**Fig. 1 | Characterization and quantitative analysis of polar V-AV pairs in designed (PTO)$_{11}$/(STO)$_6$ superlattice. a** Evolution diagram of polar V-AV pairs under thermal (T) and electric (E) fields. The purple balls represent vortices with the topological number of +1, and the orange balls represent antivortices with the topological number of -1. **b**, **c** Low-magnification MAADF-STEM image (**b**) and TEM dark-field image formed by reflection with **g**=002 (**c**), showing the high quality of the designed (PTO)$_{11}$/(STO)$_6$ superlattice and the diffraction contrast of polar vortex arrays in PTO layers. **d** An enlarged atomically resolved HAADF-STEM image for a 6-u.c. thick STO layer sandwiched between two 11-u.c. PTO layers. The polar vectors derived from the off-center displacement of Ti in the white dashed frame are depicted on the left side of **e**, revealing the existence of bound polar V-AV pairs, consistent with the simulation results on the right side. The lengths of the arrows on the rightmost panel are uniform to highlight the antivortex configuration. **f**, **g** The unit-cell-scale mappings of lattice *c* and lattice *a* extracted from **d**. **h**, **i** Corresponding simulation results of *c* and *a*. **j**, **k** The unit-cell-scale mappings of out-of-plane displacement *dz* and in-plane displacement *dx*. **l**, **m** Corresponding simulation results of *dz* and *dx*. The black rotation arrows indicate the rotation direction of polar vortices. Insets: calculation sketches of *c*, *a*, *dz* and *dx*.

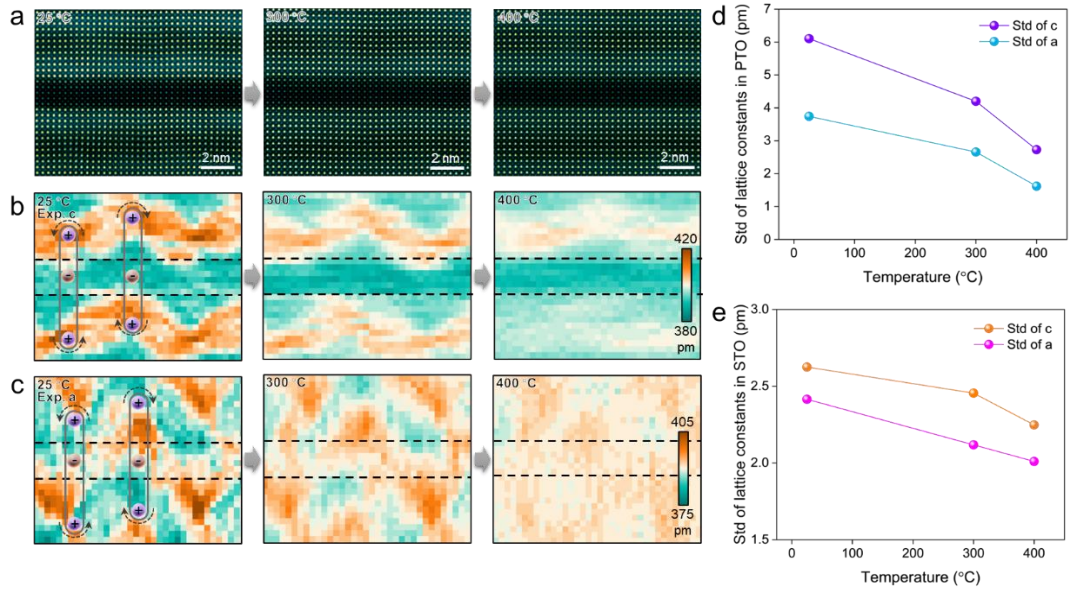

**Fig. 2 | Real-space evolution of polar V-AV pairs under thermal excitation at the atomic scale. a** A series of HAADF-STEM images acquired with increasing temperature. **b**, **c** The corresponding unit-cell-scale mapping series of out-of-plane lattice *c* (**b**) and in-plane lattice *a* (**c**) are derived from **a**, indicating the disappearance of V-AV pairs. **d**, **e** Std of lattice *c* and lattice *a* in PTO layers (**d**) and STO layers (**e**) as a function of temperature, with decreasing trends representing the lattice constant distribution from discrete to uniform.

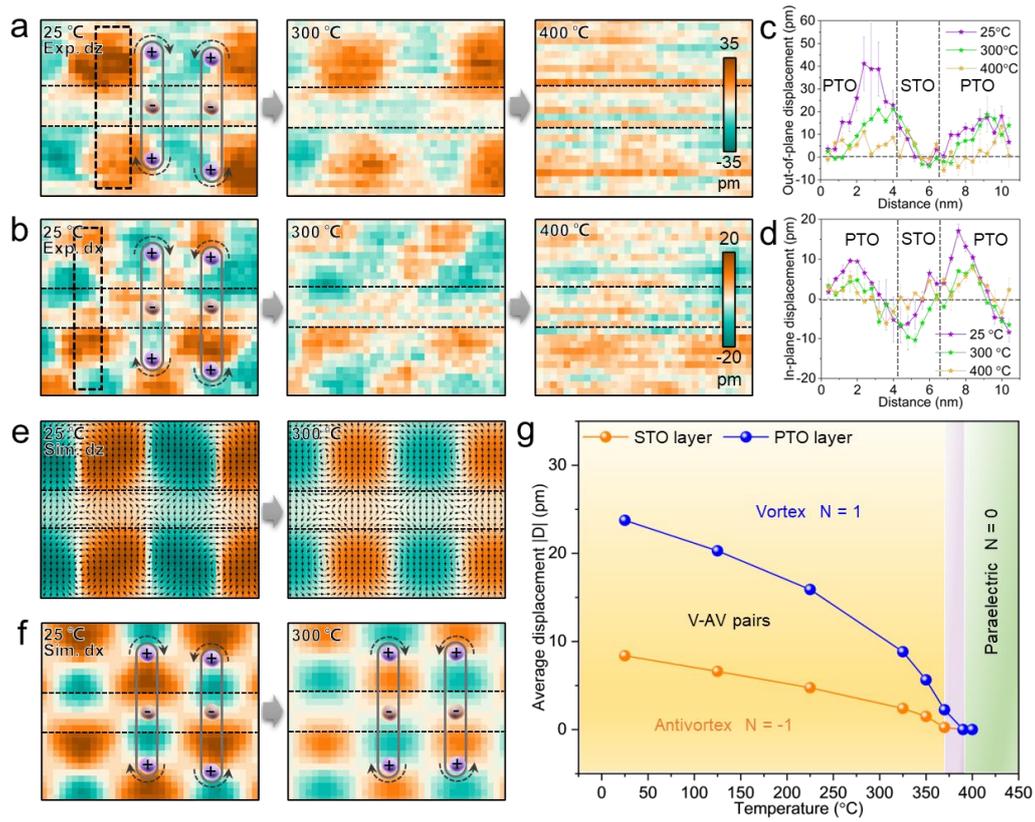

**Fig. 3 | Thermal field induced variation of polar displacement at the atomic scale.**
**a**, **b** Out-of-plane (**a**) and in-plane (**b**) displacement mapping series derived from the atomically resolved HAADF-STEM images acquired during heating. The critical Curie temperature is about 400 °C. **c**, **d** Vertical *dz* (**c**) and *dx* (**d**) line profiles corresponding to the black dashed frames in **a** and **b**, respectively. The out-of-plane and in-plane polar displacements decrease during heating, culminating with the formation of paraelectric phase. **e**, **f** Simulation results corresponding to the variation of *dz* and *dx*. **g** Temperature phase diagram of polar V-AV pairs, showing the temperature dependence of the average polar displacement |D|. N represents the topological number.

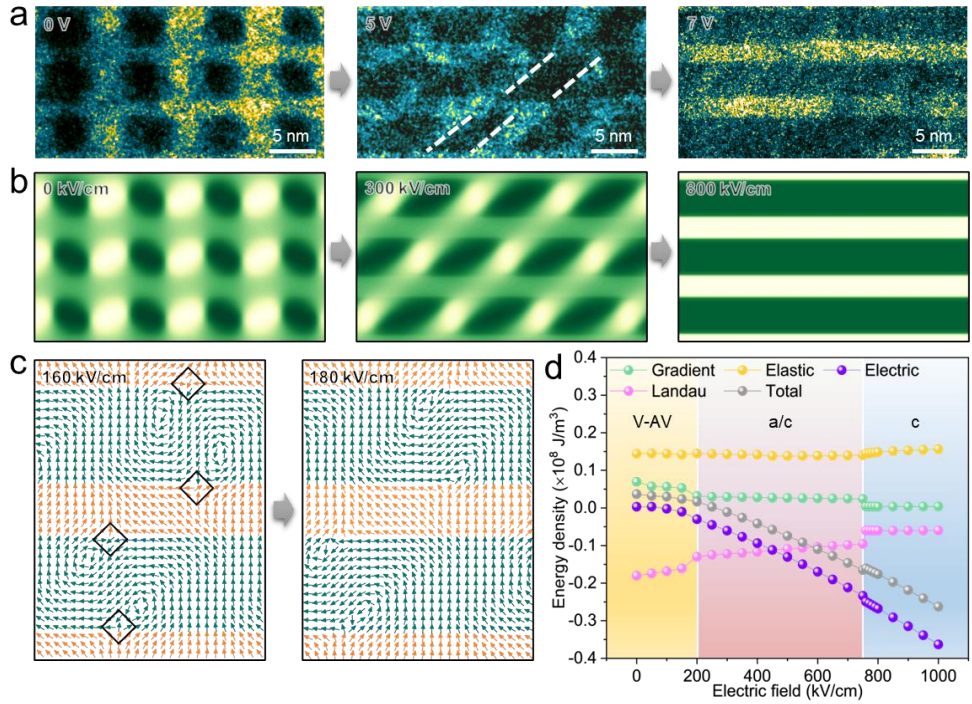

**Fig. 4 | Real-space evolution of polar V-AV pairs under electric excitation. a** Low magnification TEM dark-field image series formed by reflection with **g**=002, displaying the transformation from V-AV pairs to monodomain under applied electric bias, with a/c domain intermediate phase as outlined by the white dashed lines. **b** The corresponding phase field simulation results on the spatial distribution of out-of-plane polarization under electric fields. **c** Simulated polar maps of intermediate states captured at ~ 160 kV/cm and ~ 180 kV/cm. The black frames highlight the antivortex cores. **d** Electric field phase diagram of polar V-AV pairs, showing the evolution of energy components. Three regions can be found in which V-AV pairs, a/c domains and c domains are dominant, respectively.

# Methods

### Fabrication of superlattice

The SRO bottom electrode were epitaxial grown on (110)-DSO substrates by pulsed laser deposition (PVD-5000, with a 248 nm KrF excimer laser). The $PTO_{11}/STO_6$ superlattices were then deposited using the deposition parameters given in [58]. The PTO ferroelectric layer was deposited from a 2-inch target with 10 % excess Pb ($Pb_{1.1}TiO_3$) to compensate for Pb volatility. A nominally stoichiometric STO target was used to deposit dielectric layer STO. Alternative PTO and STO layers were deposited with the substrate held at a temperature of 600 °C and oxygen pressure of 200 mTorr. Laser energy fine tuning for the growth of the SRO, PTO and STO sublayers was crucial for the presence of polar vortex-antivortex pair in the superlattice. The sublayers' thicknesses were held by controlling the laser pulse number. After deposition, the superlattice films were cooled to room temperature at 20 °C/min under a 200-mTorr oxygen pressure.

### TEM sample preparation

Cross-sectional TEM samples for in situ thermal and electrical experiments were prepared using focused ion beam (Helios G4) etching, and transferred to processed chips (Protochips, Inc.). The sample for heating is placed on the integrated holes of the chip, while the sample for external bias is fixed on two Pt electrodes.

### Electron microscopy characterization

In situ (S)TEM experiments were performed with a Protochips double-tilt holder, and the process was observed using an aberration-corrected FEI Titan Themis G2 at an accelerating voltage of 300 kV. Dark field TEM images were acquired under the two-beam condition with **g** = 002. The convergence semi-angle for STEM imaging is 30 mrad, and the collection semi-angle snap is 50 to 200 mrad for HAADF and 39 to 200 mrad for MAADF. The atom positions determined by simultaneous fitting of two-dimensional Gaussian peaks with homemade MATLAB codes.

### Phase field simulation

In order to accurately predict the topological phase transition of vortex-antivortex pairs in superlattices, we adopt a phase field model with polarization-dependent elastic and electrostrictive coefficients[59]. The free energy density in the phase field model is as follows

$$
\begin{aligned}
f[\{\varepsilon_{ij}\}, \{E_i\}, \{P_i\}, \{P_{i,j}\}] \\
= \alpha_{ij}P_iP_j + \alpha_{ijkl}P_iP_jP_kP_l + \alpha_{ijklmn}P_iP_jP_kP_lP_mP_n + \frac{1}{2}g_{ijkl}P_{i,j}P_{k,l} \\
+ \frac{1}{2}(c_{ijkl} + a_{ijklmn}P_mP_n)\varepsilon_{ij}\varepsilon_{kl} - (q_{ijkl} + b_{ijklmn}P_mP_n)\varepsilon_{ij}P_kP_l \\
- \frac{1}{2}\kappa_0\kappa_{ij}E_iE_j - E_iP_i,
\end{aligned}
\quad (1)
$$

in which $P_i$, $\varepsilon_{ij}$ and $E_i$ are the spontaneous polarization, strain and electric field components, respectively. $\alpha_{ij}$, $\alpha_{ijkl}$ and $\alpha_{ijklmn}$ are the Landau energy coefficients, $g_{ijkl}$ are the gradient energy coefficients, $\kappa_0$ is the vacuum permittivity and $\kappa_{ij}$ are the background dielectric constants. $c_{ijkl}$ and $q_{ijkl}$ are elastic and electrostrictive coefficients of paraelectric state, respectively. In order to accurately model the elasticity and electromechanical coupling properties of ferroelectric phase, the polarization-dependent elastic coefficients $a_{ijklmn}P_mP_n$ and electrostrictive coefficients $b_{ijklmn}P_mP_n$ are added into the coefficients of elastic and electrostrictive energy densities, respectively, in Eq. (1). In which $a_{ijklmn}$ are obtained by fitting the results of experimental measurement[60], whereas $b_{ijklmn}$ are derived by using a similar approach to that of $q_{ijkl}$[61]. The polarization-dependent elastic and electrostrictive coefficients have the advantage to adjust the properties of elasticity and electromechanical coupling in ferroelectric phase automatically with the change in polarization. And these polarization-dependent coefficients will reduce to zero when the polarization disappears in paraelectric phase, see Supplementary Information and **Table S1** for details. All repeating subscripts in Eq (1) imply summation over the Cartesian coordinate components $x_i$ ($i = 1, 2$ and $3$), and ',$i$' denotes the partial derivative operator with respect to $x_i$ ($\partial/\partial x_i$).

The temporal evolution of the polarization can be described by the time-dependent Ginzburg-Landau (TDGL) equations as $\frac{\partial P_i(r,t)}{\partial t} = -L\frac{\delta F}{\delta P_i(r,t)}$, where $L$ represents the domain wall mobility, $F = \int_V f\, \mathrm{d}V$ is the total free energy in the whole simulated system, $r$ and $t$ denote the spatial position vector and time, respectively. In addition, both the mechanical equilibrium equations $\sigma_{ij,j} = \frac{\partial}{\partial x_j}\left(\frac{\partial f}{\partial \varepsilon_{ij}}\right) = 0$ and the Maxwell's equation $D_{i,i} = -\frac{\partial}{\partial x_i}\left(\frac{\partial f}{\partial E_i}\right) = 0$ are satisfied simultaneously for a body-force-free and charge-free ferroelectric system, where $\sigma_{ij}$ and $D_i$ are the stress and electric displacement components, respectively. See Supplementary Information for details of phase field simulations.

**Data availability**

Data and materials supporting this study are available from the corresponding authors upon reasonable request.

**Acknowledgements**

This research was supported by the National Natural Science Foundation of China (52125307, 11974023, 52021006, T2188101), the Key R&D Program of Guangdong Province (2018B030327001, 2018B010109009, 2020B010189001), the National Equipment Program of China (ZDYZ2015-1), and the "2011 Program" from the Peking-Tsinghua-IOP Collaborative Innovation Center of Quantum Matter. And we acknowledge the support of Guangdong Provincial Key Laboratory Program (2021B1212040001) from the Department of Science and Technology of Guangdong Province, National Program on Key Basic Research Project (2022YFB3807601) and the Key Research Project of Zhejiang Laboratory (No. 2021PE0AC02). We also thank Electron Microscopy Laboratory in Peking University for the use of the Cs-corrected electron microscope.

**Competing interests:** The authors declare no competing interests.

**Correspondence and requests for materials** should be addressed to C.T., J.L., J.W., and P. G.